\documentclass{aa}  
\usepackage{graphicx}
\usepackage[varg]{txfonts}
\usepackage{natbib}
\bibpunct{(}{)}{;}{a}{}{,} 
\usepackage{longtable}   
\begin{document}

   \title{H$\alpha$ emission-line stars in molecular clouds.}

   \subtitle{II. The M42 region.}

   \author{Bertil Pettersson
          \inst{1}
          \and
          Tina Armond\inst{2}
          \and
          Bo Reipurth\inst{3}
          }

   \institute{Observational Astronomy, Department of Physics and Astronomy, Uppsala University, Box 516,
			SE-751 20 Uppsala, Sweden\\
              \email{Bertil.Pettersson@physics.uu.se}
         \and
             Universidade Federal de Sergipe, Departamento de F\'{\i}sica, 
             Av. Marechal Rondon s/n, 49100-000, S\~ao Crist\'ov\~ao, SE, Brazil\\
             \email{tina@ufs.br}
		\and
			Institute for Astronomy and 
                        NASA Astrobiology Institute, 
                        University of Hawaii at Manoa,
                        640 North Aohoku Place, Hilo, HI 96720, USA\\
			\email{reipurth@ifa.hawaii.edu}
             }

   \date{Received 7 February 2014 / Accepted 4 June 2014 }

 
   \abstract {We present a deep survey of H$\alpha$ emission-line
     stars in the M42 region using wide-field objective prism films.
     A total of 1699 H$\alpha$ emission-line stars were identified, of
     which 1025 were previously unknown, within an area of 5\fdg5
     $\times$ 5\fdg5 centred on the Trapezium Cluster.  We present
     H$\alpha$ strength estimates, positions, and $JHK_{s}$ photometry
     extracted from 2MASS, and comparisons to previous surveys.  The
     spatial distribution of the bulk of the stars follows the
     molecular cloud as seen in CO and these stars are likely to
     belong to the very young population of stars associated with the
     Orion Nebula Cluster.  Additionally, there is a scattered
     population of H$\alpha$ emission-line stars distributed all over
     the region surveyed, which may consist partly of foreground stars
     associated with the young NGC~1980 cluster, as well as some
     foreground and background dMe or Be stars.  The present catalogue
     adds a large number of candidate low-mass young stars belonging
     to the Orion population, selected independently of their
     infrared excess or X-ray emission.}


   \maketitle
%

\section{Introduction}

The Orion Molecular Cloud is the nearest giant molecular cloud, a site
of intense star formation and one of the most thoroughly studied
regions of the sky. The cloud extends for 15 degrees in an elongated
shape that can be divided into two subregions, Orion A to the south and
Orion B to the north, each having $\sim$$10^5$ M$_{\odot}$ of molecular
gas \citep{MM86}.

The well-known OB association Orion OB1 extends through the Orion
constellation and constitutes subgroups of different ages and
locations, sometimes partially superimposed.  Associated with the OB
stars \citep{Br08}, the region contains a rich population of
intermediate and low-mass young stars.  \citet{Ba08} presents an
overview of the young stellar populations, and the morphology and
possible formation history of the cloud.
 
The prominent Orion Nebula, M42, is an HII region located in the northern
part of the Orion A molecular cloud, corresponding to Lynds 1640. To
the south, the cloud extends to Lynds 1641, also part of Orion A.

\citet{MR07} estimate a distance of 414 pc to M42, which is the
distance we assume here for the entire Orion~A cloud.

The M42 nebula is excited by the massive stars of the Orion Nebula Cluster (ONC), mainly by the
Trapezium, a tight cluster of massive young stars. The ONC is one of
the youngest and most active sites of star formation of the cloud,
studied in detail at many wavelengths (e.g. in the optical by
\citealt{Hill97} and \citealt{DR09}, in X-rays by \citealt{GF05}, in the near- and
mid-infrared by \citealt{MeGu12}).  An extensive overview of the ONC can
be found in \citet{MG08} and \citet{OM08}.

Fewer studies extend the coverage to the wider areas outside of the ONC.
\citet{CH01} used 2MASS to identify over a thousand young stars in Orion A
through their variability. They examined a 0\fdg84 $\times$
6\degr\ strip centred on the ONC and found a distribution that correlates
with the H$\alpha$ emission-line stars and the molecular gas. Most of
the variability can be explained by cold and hot starspots, by accretion, 
or by varying extinction.

\citet{DF09} imaged Orion A in H$_2$~2.122~$\mu$m in search of
molecular hydrogen outflows and their sources, which they classify as
mostly protostellar.
The {\it Spitzer} survey from \citet{MeGu12} covers Orion A and B and
classify 2991 pre-main-sequence stars with disks and 488 as likely
protostars, based on mid-infrared colours that indicate the presence of
a dusty envelope or circumstellar material.  A catalogue with five
mid-infrared colours is presented.  They detect variability in 50\% of
the young sources.

The powerful radiation field and the expanding HII region of the ONC
have cleared away some of the cloud surrounding the cluster, making
part of the young stellar content visible at optical wavelengths. Such
visible young stars are recognisable through the H$\alpha$ emission
line which is caused by ongoing accretion processes.

Large-scale objective prism surveys focused on the H$\alpha$ line have
for a long time been an important way to identify young stars in
clusters and throughout molecular clouds. In Orion, the pioneering
work of \citet{Haro53} identified 255 H$\alpha$ emission-line stars in
an area of 3.5 square degrees around the brightest part of M42.
\citet{PaCh82} added additional stars to Haro's list, following his
nomenclature, and listed a total of 534 stars in a 5 square degree
region.  Additional information on many of these stars is summarised
in the catalogues of \citet{HR72} and \citet{HB88}, which were for
years a reference for many of the optically visible young stars in Orion.

Subsequently, a large and systematic H$\alpha$ emission survey was
performed throughout the Orion constellation using the Kiso Schmidt
telescope, covering an area of 300 square degrees, resulting in about
1200 emission-line stars detected with limiting magnitude of V=17.5.
The results were published in a series of papers, each covering one
region in Orion \citep{WKY89,KYW89,WKY91,WKY93,NW95}.  Their
coordinate range in the sky is about $4\fh8 < \alpha < 6\fh2$,
$-13\degr < \delta < +7\degr$ (J2000).

\citet{BW92} present an overview of the low mass star formation in
Orion, including a comprehensive list of all H$\alpha$ emission-line
stars compiled from all the surveys known until then, including their
own work in the L1641 cloud \citep{WB92}.  The catalogue contains 1297
stars in the range $5\fh45 < \alpha < 5\fh81$, $-10\degr < \delta <
+02\fdg6$ (J2000).  It includes all the Kiso surveys, except for the
last two published in 1993 and 1995 (114 stars).

Recently, \citet{SE13} performed a survey for H$\alpha$ emission-line
stars in an area of one square degree around the ONC, using slitless
grism spectroscopy. They detected 587 stars with emission, of which 99
are new findings.

Another technique used to detect H$\alpha$ emission is through the use of
narrow and wide filters centred on the H$\alpha$ line. This technique
was used by Da Rio et al. (2009). In a 30$'$$\times$30$'$ region
approximately centred on the ONC they detected 323 stars with a
strong line excess exceeding 50~\AA\ in equivalent width and 315 stars
with weak H$\alpha$ excess corresponding to equivalent widths in the 5 $-$
 50 \AA\ range. As noted by the authors, the uncertainty in the
derived H$\alpha$ emission is difficult to evaluate, and may be
affected by various problems, including the non-uniformity of the strong
nebular H$\alpha$ emission. Their survey is mainly focused on the
central region of the ONC where we are most affected by the strong
nebular background, and hence there is little overlap between our
surveys. Finally, one can of course detect H$\alpha$ emission by obtaining
spectra of individual stars. With modern multi-slit spectrographs this
is becoming feasible, as for example demonstrated by the kinematic
study of \citet{FH08}, who observed 1215 stars in the ONC,
of which 1111 stars were confirmed as members. The disadvantage of
this method from the point of view of searching for new young stars is
that the targets first need to be selected by some criteria, so the search
is not unbiased. In the following, we compare our results only with other
objective-prism or grism surveys.

In this paper we continue to present the results of a series of large
scale searches for H$\alpha$ emission-line stars using photographic
films and the large objective prism at the ESO Schmidt telescope, before it
ceased operations in 1998. The first of the series \citep[from now on
Paper~I]{RPA04} studied the NGC 2264 star forming region, where we
detected 357 H$\alpha$ emission-line stars, of which 244 were new
findings. In the present paper we focus on Orion.
\defcitealias{RPA04}{Paper~I}

The area surveyed here is 5\fdg5 $\times$ 5\fdg5 on the sky centred
on the Orion Nebula Cluster, covering a wider area than all of the
previous surveys except the Kiso survey, but deeper than that one.  By
going deeper, we are sensitive to lower-mass stars, and also to stars
suffering higher extinction.

H$\alpha$ emission in young stars is widely assumed to be triggered by
accretion from a circumstellar disk through funnel flows onto the
star. H$\alpha$ emission is thus a measure of a temporary condition,
and is therefore expected to be variable, although the timescale of
variability is poorly known. An advantage of using H$\alpha$ emission
to identify young stars is that even stars with little circumstellar
material, which are thus difficult to identify as young in infrared
surveys, can accrete and produce the tell-tale H$\alpha$ emission.
H$\alpha$ emission-line surveys and infrared surveys therefore to some
extent complement each other.


\section{Observations}

Spectral films (sensitised Kodak 4415) were taken between 1995 and
1997 at the ESO Schmidt telescope at La Silla, equipped with an
objective prism that yields a dispersion of 800 \mbox{\AA\ mm$^{-1}$}
at H$\alpha$, as already described in detail in Paper~I.  The RG 630
filter used provides a spectral range from 630 nm to 690 nm, a narrow
range centred on the H$\alpha$ line.  Exposure times ranging from 15
minutes to 150 minutes allowed the detection of emission in both faint
and bright stars. The very central region of M42, where the nebulosity is too
intense, was impossible to examine, even in the shortest exposures.
The size of the films (30 cm $\times$ 30 cm) corresponds to a field of
5\fdg5$ \times$ 5\fdg5 on the sky. All exposures were centred on the position
$\alpha$:~5$^\mathrm{h}$35$^\mathrm{m}$ $\delta$:~$-5$\degr 25\arcmin\
(J2000).  Table \ref{tbl1} lists the films employed in the present
study.

\begin{table}
\caption{List of Schmidt films }             
\label{tbl1}      
\centering                          
\begin{tabular}{c c c c c}        
\hline\hline                 
Date &  No. & Exp. time & Filter & Seeing \\ 
     &      & (minutes) &        & (\arcsec) \\    
\hline                        
1995 Nov 24 & 12093B & 120 & RG630 & 1.0  \\
1995 Nov 27 & 12103B &  15 & RG630 & 1.0  \\
1996 Feb 13 & 12175B & 120 & RG630 & 0.9  \\
1996 Nov 20 & 12852B &  40 & RG630 & 0.5  \\
1996 Dec 13 & 12896B & 150 & RG630 & 0.8  \\
1996 Dec 14 & 12898B & 150 & RG630 & 0.7  \\
1997 Jan 06 & 12935B &  40 & RG630 & 0.5  \\
1997 Jan 11 & 12952B &  90 & RG630 & 0.85 \\
1997 Jan 31 & 12972B &  90 & RG630 & 0.8  \\
\hline                                   
\end{tabular}
\end{table}


\section{The survey}

The Schmidt films were visually inspected with the use of a binocular
microscope in search of emission against the continuum of the stars.
About 2360 candidate H$\alpha$ emission-line objects were initially identified.
An H$\alpha$ strength was assigned to each object, in a range from 1
to 5, following the same procedure used for our NGC 2264 survey
\citepalias{RPA04}.  The H$\alpha$ strength is defined so that 1 is
weak emission against a strong continuum and 5 is strong emission
against a weak or absent continuum.  Only the stars identified as
unequivocally possessing emission were retained, and emissions identified
as coming from HH objects and some galaxies were also removed, which altogether reduced
the number of stars in our survey to 1699.

The POSS-II surveys (from 1987/1988 Digitised Sky Survey, DSS) that provide 
$m_{J}m_{F}m_{N}$ images with a plate scale of \mbox{1\arcsec pixel$^{-1}$}, 
were used to obtain initial coordinates for the stars. 
The images were retrieved in {\tt FITS} format and had astrometric 
information in their headers. The task {\tt MAKEWCS} from the {\tt IRAF} 
image reducer package was used on each image to add a coordinate system 
that can be understood by {\tt IRAF}. 
The {\tt IRAF} task {\tt IMEXAMINE} was used to determine the
coordinates of each star visually identified in the finding charts.
$R$ images were used to measure the coordinates for most of the stars, 
while $I$ images were used in regions where the nebulosity is particularly bright 
at optical wavelengths. 

\begin{figure}
   \centering
   \includegraphics[width=\columnwidth]{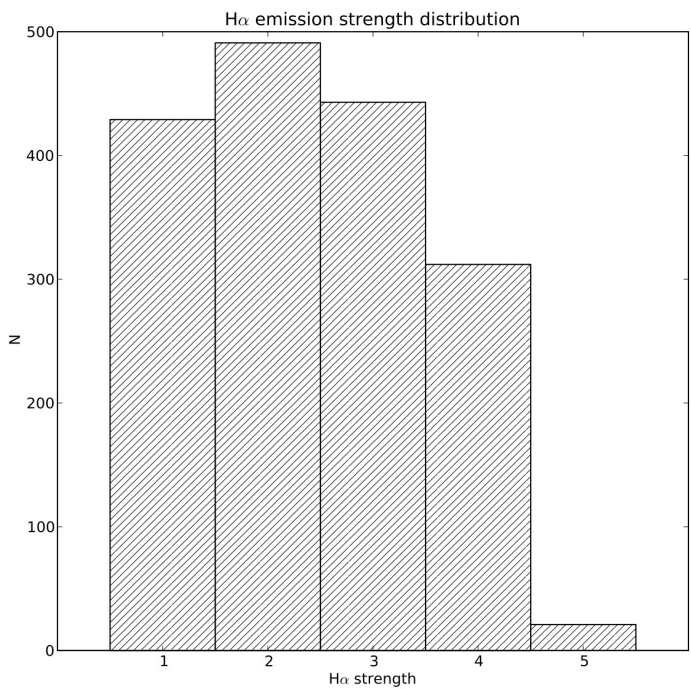}
      \caption{Distribution of H$\alpha$ emission strength assigned 
by eye to the stars. When there was a range of values, the lower one was used. }
         \label{fig-dist}
   \end{figure}

These initial coordinates were searched (within a radius of 2\arcsec)
among the stars of the 2 Micron All Sky Survey (2MASS) catalogue in order 
to obtain near infrared $JHK_s$ magnitudes. The 2MASS coordinates have 
expected uncertainties of 0\farcs1, better than what we achieved with
the DSS images. Therefore, the present survey uses the 2MASS coordinates.
In some cases with close pairs the 2MASS listed coordinates were not 
resolved and coordinates were visually measured from 2MASS J frames.
Finding charts for the new stars, extracted from the DSS, are presented
in Figs.\ 8-18, available in the electronic version.
\rm
Emission-line strengths were estimated by eye on each film separately.
Differences in estimates are indicated as hyphenated values, and may
represent either intrinsic variability and/or uncertainty in the estimate.
Nearly 18\% of the stars present such differences. The distribution of
H$\alpha$ emission strength is shown in Fig.\ \ref{fig-dist}.

The USNO-B catalogue was used to get blue, red, and infrared
photographic magnitudes ($m_J$, $m_F$, and $m_N$)
for most of the stars.  
Blue and red magnitudes were
extracted from the GSC-2.2 catalogue for a few additional stars that
were only detected in that catalogue.  Both catalogues provide optical
magnitudes for about 80\% of our stars.  We also got
crossidentifications with the 2MASS
catalogue and were able to obtain near-infrared $JHK_s$ magnitudes for
all but eight stars.

\begin{figure}
   \centering
   \resizebox{\hsize}{!}{\includegraphics{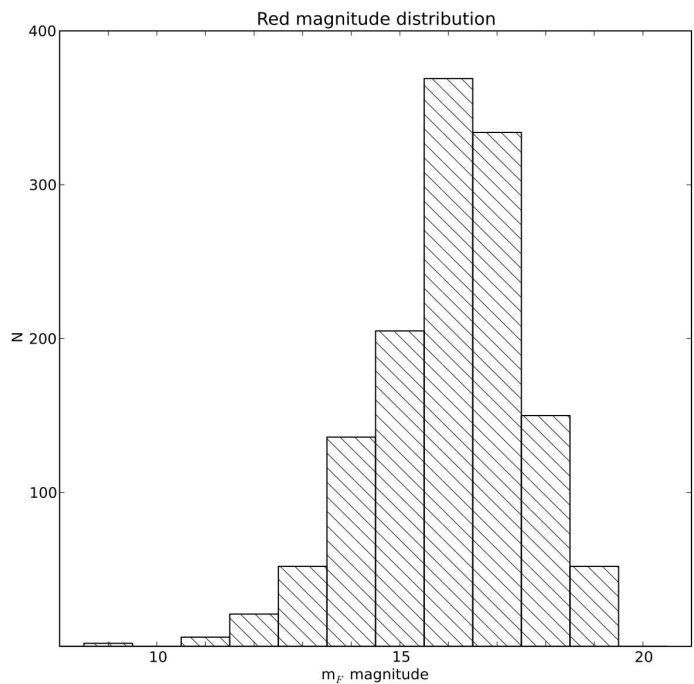}}
      \caption{Brightness distribution of the 1327 H$\alpha$ emission stars with
a $m_F$ magnitude. The limiting magnitude of our survey is about 17-18 mag.
}
         \label{fig-hist}
   \end{figure}

   The brightness distribution of the 1327 H$\alpha$ emission-line
   stars with a (red) $m_F$ magnitude is shown in Fig.\ \ref{fig-hist}. 
   The limiting magnitude of our survey is about
   $m_F$=17-18 mag.  The peak at 16 mag indicates that the vast
   majority of the stars detected in our survey are low-mass objects,
   most likely T~Tauri stars. The faintest stars either have higher
   extinction or they are brown dwarfs.

\begin{figure}
   \centering
   \includegraphics[width=\columnwidth]{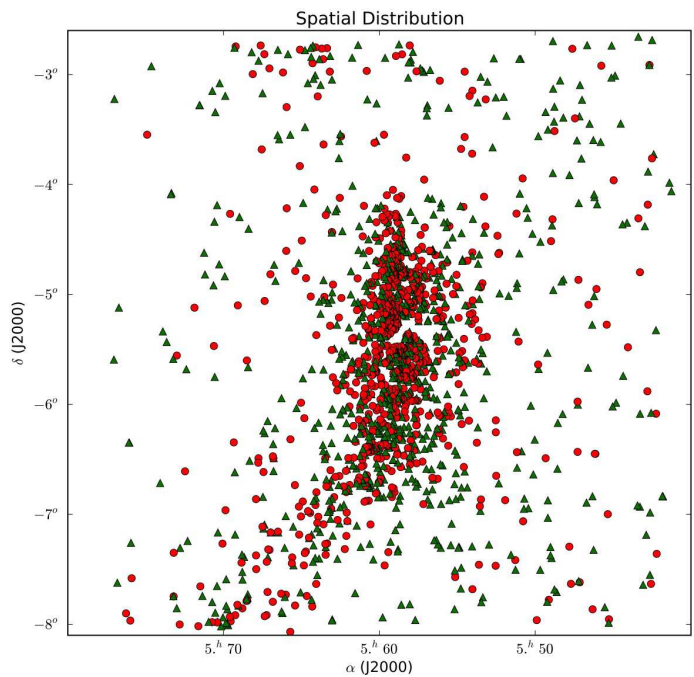}
      \caption{Spatial distribution of the 1699 H$\alpha$ emission stars in
our M42 survey. The patches that are empty of stars in the central region are
located where the HII region is very intense, complicating the detection
of H$\alpha$ emission-line stars. Stars with an H$\alpha$ value of 1-2 are
plotted with green triangles and 3-5 as red dots. See text for a discussion.}
         \label{fig-map}
   \end{figure}

   The spatial distribution of the emission-line stars is shown in
   Fig.\ \ref{fig-map}.  
   The patches that are empty of stars in the central
   region are located where the nebulosity is too intense, even in the
   shortest exposures.

\begin{table*}
\caption{H$\alpha$ emission-line stars in M 42}
\label{tbl2}
\centering
\renewcommand{\footnoterule}{}  

{\scriptsize
\tabcolsep 4pt
\begin{tabular}{lllllllcccccccccl}     
\hline\hline
ESOH$\alpha$ & GCVS & KisoH$\alpha$\tablefootmark{a} & HBC\tablefootmark{b} & Haro\tablefootmark{c} & PaCh & Other\tablefootmark{d}  &
$\alpha$(2000)\tablefootmark{e} & $\delta$(2000)$^e$ & H$\alpha$\tablefootmark{f} &  
$m_{J}$\tablefootmark{g} & $m_{F}^\mathrm{g}$ & $m_{N}^\mathrm{g}$ & $J$\tablefootmark{h} & $H^\mathrm{h}$ & $K_s^\mathrm{h}$&Notes
\tablefootmark{i} \\
\hline
	&	V731	&	76-59	&		&	4-48	&	70	&		&	05	33	47.71	&	$-$04	52	08.5	&	2	&	16.05	&	14.54	&	13.01	&	11.64	&	11.03	&	10.73	&		\\
	&		&	76-61	&		&	4-99	&	73	&		&	05	33	48.22	&	$-$05	13	26.2	&	3	&	18.36	&	16.50	&	14.90	&	12.99	&	12.14	&	11.71	&		\\
857	&		&		&		&		&		&	*	&	05	33	48.34	&	$-$05	22	39.4	&	1	&		&		&	13.96	&	12.70	&	11.99	&	11.72	&		\\
858	&		&		&		&		&		&		&	05	33	48.51	&	$-$07	13	59.3	&	2	&	12.63	&	18.72	&	15.83	&	13.81	&	13.20	&	12.91	&		\\
	&	V354	&	76-63	&		&	4-300	&	78	&		&	05	33	49.54	&	$-$05	36	20.8	&	2	&		&		&	15.15	&	12.37	&	11.32	&	10.71	&		\\
	&	HX	&	75-186	&		&	4-18	&	75	&		&	05	33	50.27	&	$-$04	38	34.2	&	3	&	14.61	&	13.35	&	12.30	&	11.21	&	10.52	&	10.29	&	11	\\
859	&		&		&		&		&		&	*	&	05	33	50.74	&	$-$05	00	39.5	&	2	&	18.01	&	16.59	&	14.91	&	13.18	&	12.42	&	12.12	&		\\
	&		&	76-64	&		&	4-39	&	77	&		&	05	33	51.35	&	$-$04	48	22.2	&	1-2	&	16.26	&	14.68	&	13.64	&	12.48	&	11.61	&	11.27	&		\\
	&	HY	&	76-66	&		&	4-159	&	81	&		&	05	33	52.36	&	$-$05	41	50.2	&	1	&	15.60	&	14.52	&	12.82	&	11.72	&	10.90	&	10.56	&		\\
860	&		&		&		&		&		&		&	05	33	52.62	&	$-$04	57	51.0	&	3	&	18.67	&	16.91	&	15.43	&	13.37	&	12.74	&	12.46	&		\\
861	&		&		&		&		&		&		&	05	33	53.38	&	$-$07	14	11.6	&	2	&	17.58	&	16.41	&	13.81	&	12.88	&	12.30	&	12.06	&	2,8	\\
\hline
\end{tabular}
}  
\\
\tablefoot{
The full table with all the 1699 H$\alpha$ emission-line stars is available 
electronically. A few lines are reproduced here only for guidance regarding 
format and content.
\tablefoottext{a}{Kiso H$\alpha$ catalogue from \citet{WKY93}.}
\tablefoottext{b}{\citet{HB88} catalogue.}
\tablefoottext{c}{\citet{Haro53} catalogue: Haro\,4-1 to 4-255, \citet{PaCh82}: Haro\ 4-256 to 4-495, \citet{HM53}: Haro\ 5-1 to 5-98.}
\tablefoottext{d}{An asterisk means a star also listed in \citet{SE13}}
\tablefoottext{e}{Positions extracted from the 2MASS All-Sky Catalog.}
\tablefoottext{f}{The H$\alpha$ strength is defined so 1 is weak emission against 
a strong continuum and 5 is strong emission against a weak or invisible continuum. Hyphenated 
values may represent either variability and/or uncertainty in the estimate. A '+' indicates resolved spectra but unresolved DSS image.}
\tablefoottext{g}{The magnitudes $m_J$, $m_F$, and $m_N$ are from the blue
(IIIaJ emulsion), red (IIIaF emulsion), and infrared (IV-N emulsion)
digitised sky surveys extracted from USNO-B catalogue or from GSC 2.2.}
\tablefoottext{h}{$JHK_s$ magnitudes extracted from the 2MASS All-Sky Catalog.}
\tablefoottext{i}{Notes to individual stars.}
}   
\end{table*}

\subsection{Comparison with previous surveys}

There were 674 stars in our survey that had  already been detected 
in previous objective prism H$\alpha$ surveys,
yielding a total of 1025 new H$\alpha$ emission-line stars in the
Orion region. Of the previously known emission-line stars, 436 are found in the
compilation from \citet{BW92}, which comprises most of the work done
until then in surveying the Orion region for H$\alpha$ emission-line
stars. The star Strom~6 appears in \citet{BW92} but is designated as L1641 N.
In addition, 238 stars were found to be in common with the recent survey of \citet{SE13}.
Since the stars of \citeauthor{SE13} were not given an identifier and their survey
was done simultaneously with our present survey, we have retained
our ESO-H$\alpha$ numbers for those stars that we have in common.

The \citet{HB88} catalogue (HBC) has 126 stars inside our field, of
which we detected 55 (44\%).  The HBC catalogue was compiled from
stars observed with different techniques and at different epochs. At
least a dozen HBC stars are located in the brightest M42 area, where
our films are overexposed.  

Haro's catalogue \citep{Haro53,PaCh82} has 530 stars in our field, of
which 359 were detected by us (68\%).  Their coordinates are in some
cases rather uncertain, and this plus variability in the line emission
must account for the 32\% of stars not detected by us.

Among the 471 stars from the Kiso H$\alpha$ catalogue of \citet{WKY93}
that lie inside our field, 314 were also detected by us, a fraction of
about 67\%.  The Kiso catalogue was constructed in a more uniform way,
similar to our survey, but the ratio of stars in common is still low.
Our survey is much deeper than Kiso's; our long exposures reach 150
minutes while Kiso's maximum exposures are 90 minutes, and in the same
area surveyed they detected a quarter of the number of
emission-line stars detected in our survey.  In general the H$\alpha$
emission strengths assigned by eye (with the same scale) in both
surveys agree very well, but it is noteworthy that most of the stars
in common have larger emission strengths, peaking at 3.  The
undetected Kiso objects were all re-examined on our films; some were
located in the overexposed areas, while others were classified by us
as possible emitters, but were not included in our final table.  About
60\% of the undetected Kiso objects certainly show no emission at our
resolution.  
\begin{figure}
   \centering
   \includegraphics[width=\columnwidth]{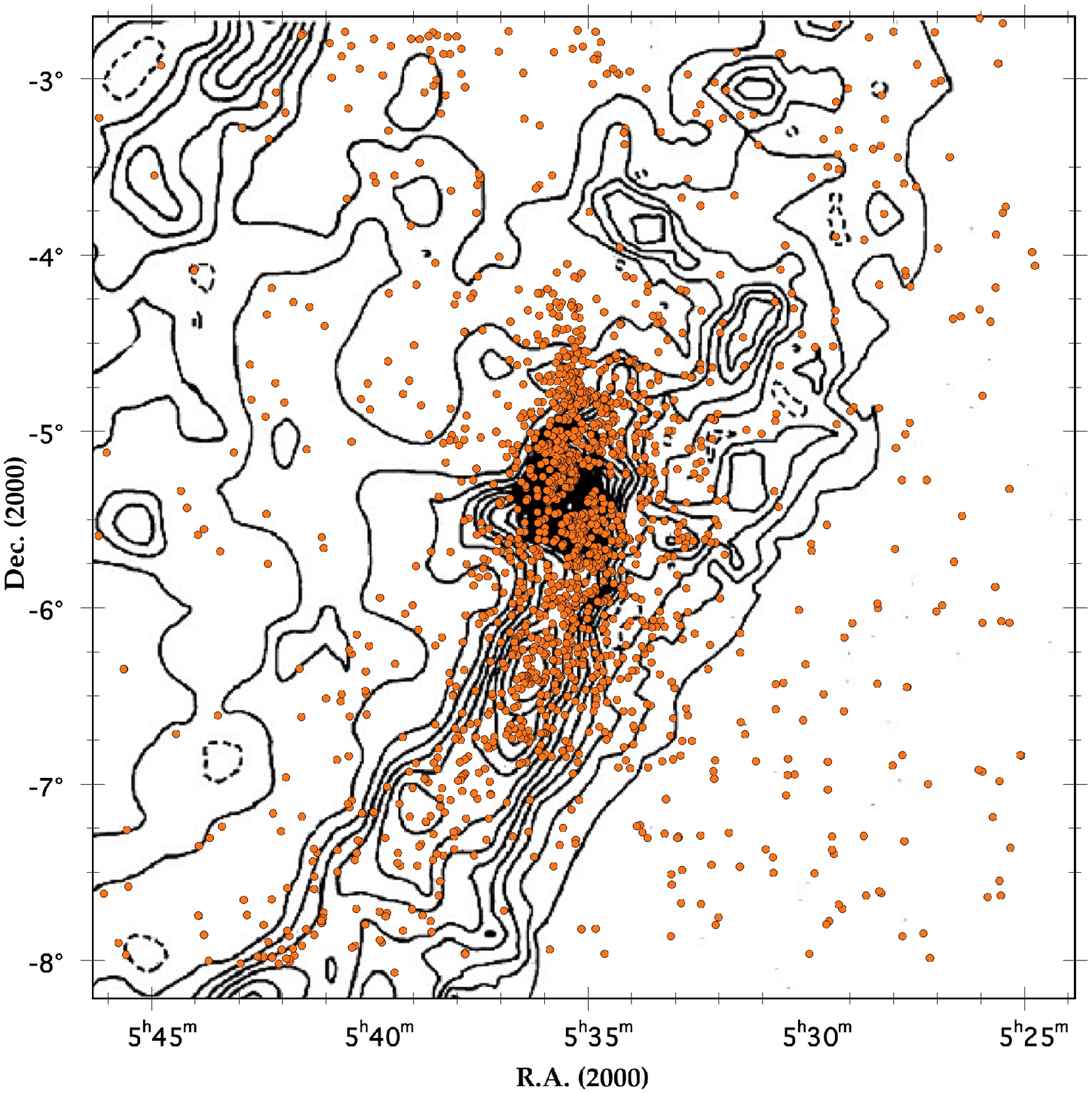}
      \caption{ Distribution of the H$\alpha$ emission stars
(orange dots) overplotted on a CO map contour from \citet{MM86}.
}
         \label{fig-co}
   \end{figure}

\subsection{Description of the table}

A table with the H$\alpha$ emission stars in M42 was built using the 
same criteria as the NGC 2264 survey \citepalias{RPA04}. Some representative lines
are listed in Table \ref{tbl2}.  The full table is available in the
electronic version, and will also be available electronically from
CDS.  The first columns are: the ESO H$\alpha$ identification number,
assigned only to the stars not detected previously, the variable name or number
from General catalogue of variable stars \citep{Kuka85}, followed by the
identification numbers of the H$\alpha$ emission surveys from
\citet{WKY93}, \citet{HB88}, \citet{Haro53}, \citet{PaCh82}, and other
possible names.
The coordinates (J2000) are given in the next columns,
followed by the H$\alpha$ emission strength and the magnitudes obtained in the
USNO-B catalogue ($m_J$, $m_F$, and $m_N$) and in 2MASS ($JHK_s$).
The last column provides comments on individual objects. Finding charts 
for individual stars, often located in crowded
    regions or as components in close binaries, are presented in
    Figures~8--18, which are available in the electronic version
    only.

\subsection {Spatial distribution and relation to CO clouds}

The main concentration of H$\alpha$-emitting stars corresponds very
well to the distribution of the gas in the Orion~A Cloud, as shown in
Fig.\ \ref{fig-co} 
in a comparison with a CO map from \citet{MM86}.
The region with the largest number of stars matches the area with
strongest CO emission, and the stellar density continues to trace the
south-eastern L1641 dark cloud. The vast majority of these stars are
likely to represent young T~Tauri stars with ages approximately
similar to the age of the ONC.

\begin{figure}
   \centering
   \includegraphics[width=\columnwidth]{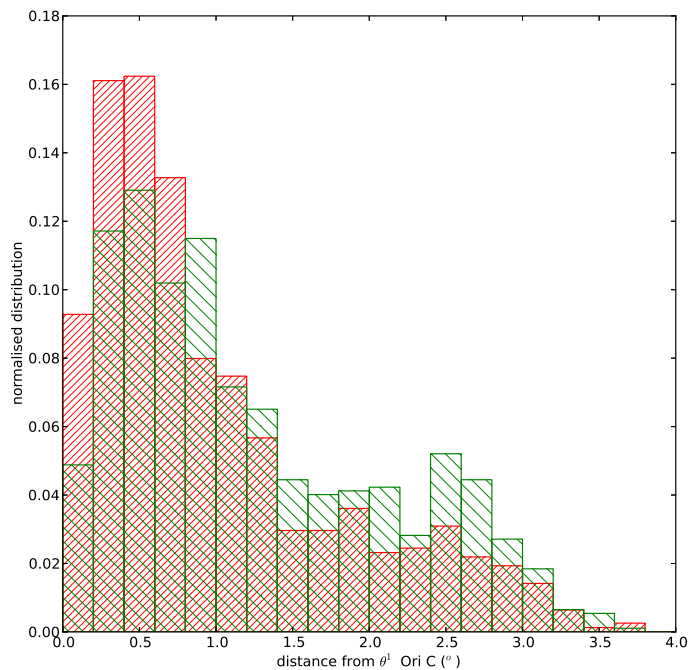}
      \caption{Distribution of stellar density with increasing distance 
      to $\Theta$ Ori. A difference in distribution between
    stronger-lined stars (red, dense hatch) and weaker-lined stars (green, sparse hatch) is
    apparent, with the former more clustered around the Trapezium and
    a wider distribution of the latter.}
         \label{fig-dens}
   \end{figure}

Additionally, there is an almost uniform distribution of stars all
over the 30 sq. degree area of our survey.  This
  population may have a more complex composition. First, some of the
  stars may originate in the ONC, and have been scattered away as a result of
  N-body interactions in the cluster. Second, there is a slightly
  higher density of H$\alpha$ emitters to the west of the cloud, where
  the extinction is quite a bit lower than to the east.  This might
  indicate that some of the stars could be background stars, for
  example distant Be stars. Third, and perhaps most important, it is
  known that there are several generations and populations of young
  stars in the general direction of the ONC (e.g. \citealt{Blaa64},
  \citealt{WaHe78}, and \citealt{GoLa98}). Most recently,
  \citet{BoAl14} have argued that there is a rich population in
  front of the Orion~A cloud, centred on and probably originating in
  the little-studied cluster NGC~1980 just south of the ONC. At an age
  of $\sim$4--5~Myr, this foreground cluster would still have many
  low-mass members with H$\alpha$ emission.
  We would expect a younger
  population to have generally stronger
    emission, whereas an older population would have fewer stars with
    strong emission. We have examined this, and in Fig.\ \ref{fig-dens} 
    the
    stellar density of the two groups is plotted as a function of
    distance to the centre of the ONC. It is evident that the
    stronger-line stars (H$\alpha$-strength 3,4,5) are indeed more
    concentrated towards the centre of the ONC, whereas the
    weaker-line stars (H$\alpha$-strength 1,2) are more
    distributed. This supports the idea of \citet{BoAl14} that we
    are seeing the mixture of two populations of young stars, one
    related to the ONC and another to the slightly older NGC 1980
    region. Additional interlopers from, for example, the NGC~2024 and
    $\sigma$~Ori regions may be present as well.

  It thus appears that the H$\alpha$ emission stars found across our
  field represent two distinct groups, one related to the young
  population of ONC stars, and another group more uniformly spread
  across the field with a mixed origin, some coming from the
  foreground NGC~1980 cluster and others representing scattered ONC
  stars or background Be stars.

\subsection {Near-infrared properties}

\begin{figure}
   \centering
   \includegraphics[width=\columnwidth]{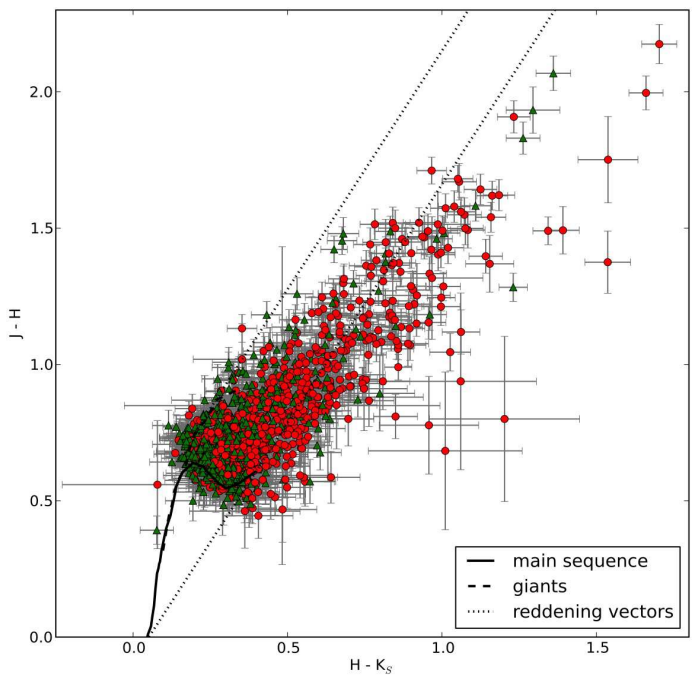}
   \caption{Near-infrared colour-colour diagram based on 2MASS data
     showing all the H$\alpha$ emitters in M42 with 2MASS detections
     in all the three bands ($JHK_s$), except upper limits.
     Main-sequence and giant loci from \citet{BB88}, corrected to the
     2MASS photometric system \citep{CH01} and the extinction law from
     \citet{RL85}, are represented in the figure. Symbols as defined in Fig. 3.
      }
         \label{fig-coco}
   \end{figure}

The H$\alpha$ emission stars of our survey that are detected in all
three bands of the 2MASS near-infrared catalogue are plotted in a
colour-colour diagram ($J-H$) vs.\ ($H-K_s$) in Fig.\ \ref{fig-coco}.
Also plotted are the locations of main-sequence and giant stars from
\citet{BB88} corrected to the 2MASS photometric system \citep{CH01},
with interstellar reddening vectors from \citet{RL85}.  

Of 1664 stars with valid $JHK_s$ magnitudes, 261$\pm$51 have
infrared excess, one indication of the presence of circumstellar
material. 
The uncertainties in the 2MASS magnitudes are on average
0.04 mag and can raise or lower the number of infrared-excess stars by
about a quarter.  

The fraction of H$\alpha$ emitting stars with infrared excess is only
$\sim$16$\pm$4\%.  In our survey of NGC~2264 (Paper~I) performed with the same
equipment, we found that 23$\pm$8\% of the H$\alpha$ emitters show an
infrared excess, which is similar within the cited errors.  There can
be several reasons that only a relatively small fraction of H$\alpha$
emitters have infrared excess. First, many weak-line T~Tauri stars
have little circumstellar material. Second, not all H$\alpha$ emitting
stars are necessarily young, for example dMe stars and Be stars will
have H$\alpha$ emission, but no infrared excess.  Third, although the
population of stars from the foreground cluster NGC~1980 is young, at
an age of 4--5~Myr there will be many fewer classical T~Tauri stars
than among the ONC population. Fourth, even young stars with little
circumstellar material can from time to time accrete and thus
temporarily produce H$\alpha$ emission.

A small number of stars (less than 2\%) are located in the
so-called forbidden region to the left of the main-sequence locus and the
reddening vectors, but all of them are faint and/or located in
nebulous regions, suggesting that their 2MASS uncertainties are
underestimated.

\begin{figure}
   \centering
   \includegraphics[width=\columnwidth]{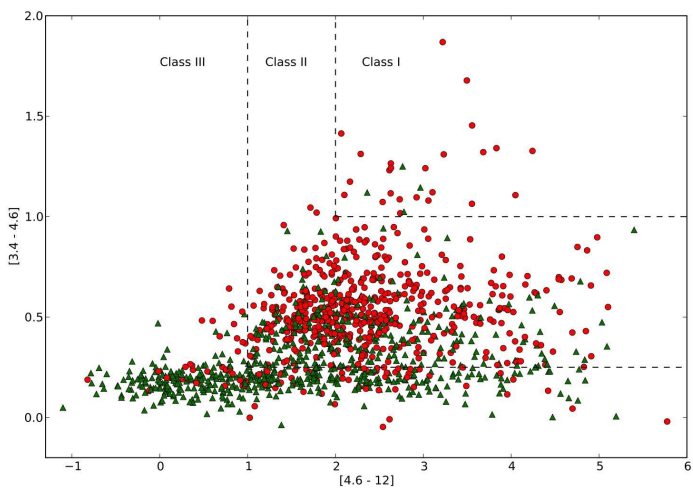}
      \caption{Distribution in a {\em
    WISE} two-colour diagram, separating stronger-lined stars (red dots)
    from weaker-lined stars (green triangles). The dashed lines separate
    Class~I, II, and III sources, see discussion in \citet{KoLe12}
    }
         \label{fig-wise}
   \end{figure}

In Fig.\ \ref{fig-wise} 
    the distribution of the stronger-lined
    stars (H$\alpha$-strength 3,4,5) and the weaker-lined stars
    (H$\alpha$-strength 1,2) is plotted in a {\em WISE} two-colour diagram used
    to classify young stars into Classes I, II, and III based on
    infrared excess as an indicator for the presence of circumstellar
    material (e.g. \citealt{KoLe12}). Of the few stars that fall in
    the Class~I category almost all have strong H$\alpha$
    emission. The Class~II category is also dominated by stars with
    stronger emission, although a sizeable number of weaker-lined stars
    are also in this category. The Class~III category is strongly
    dominated by stars with weak line-emission. Since
    H$\alpha$-emission is mainly an indicator for ongoing accretion,
    these results follow the intuition that stars with more
    circumstellar material are more likely to be actively accreting.
    
\section{Conclusions}

We have observed the central Orion region in a deep wide-field survey
for H$\alpha$ emission-line stars, detecting 1699 stars with emission,
of which 1025 were previously not
known to show H$\alpha$ emission. The stars fall into two groups,
one that contains most of the stars and is distributed along the
L1641 cloud, and another that is more or less uniformly distributed.
A detailed photometric and spectroscopic follow-up study is required
to more precisely identify which stars are likely to be members of the
young ONC population, or of the slightly less young foreground NGC~1980
population, or are general foreground or background stars.


\begin{acknowledgements}

We are grateful to Guido and Oscar Pizarro, who as telescope
 operators obtained all the films used in this survey, and to the
 referee, Herv\'e Bouy, for a very helpful report.  TA acknowledges
 financial support from CNPq-Brazil under the process number
 200279/2009-2.  BR acknowledges support through the NASA
 Astrobiology Institute under Cooperative Agreement No. NNA09DA77A.
 We acknowledge use of the Digitised Sky Survey and of the SIMBAD
 database, operated at CDS, Strasbourg, France, and of NASA's
 Astrophysics Data System Bibliographic Services.  This publication
 makes use of data products from the Wide-field Infrared Survey
 Explorer and from the Two Micron All Sky Survey.
 
\end{acknowledgements}


   \onlfig{
   
   \begin{figure*}
   \centering
   \includegraphics[width=16.8cm]{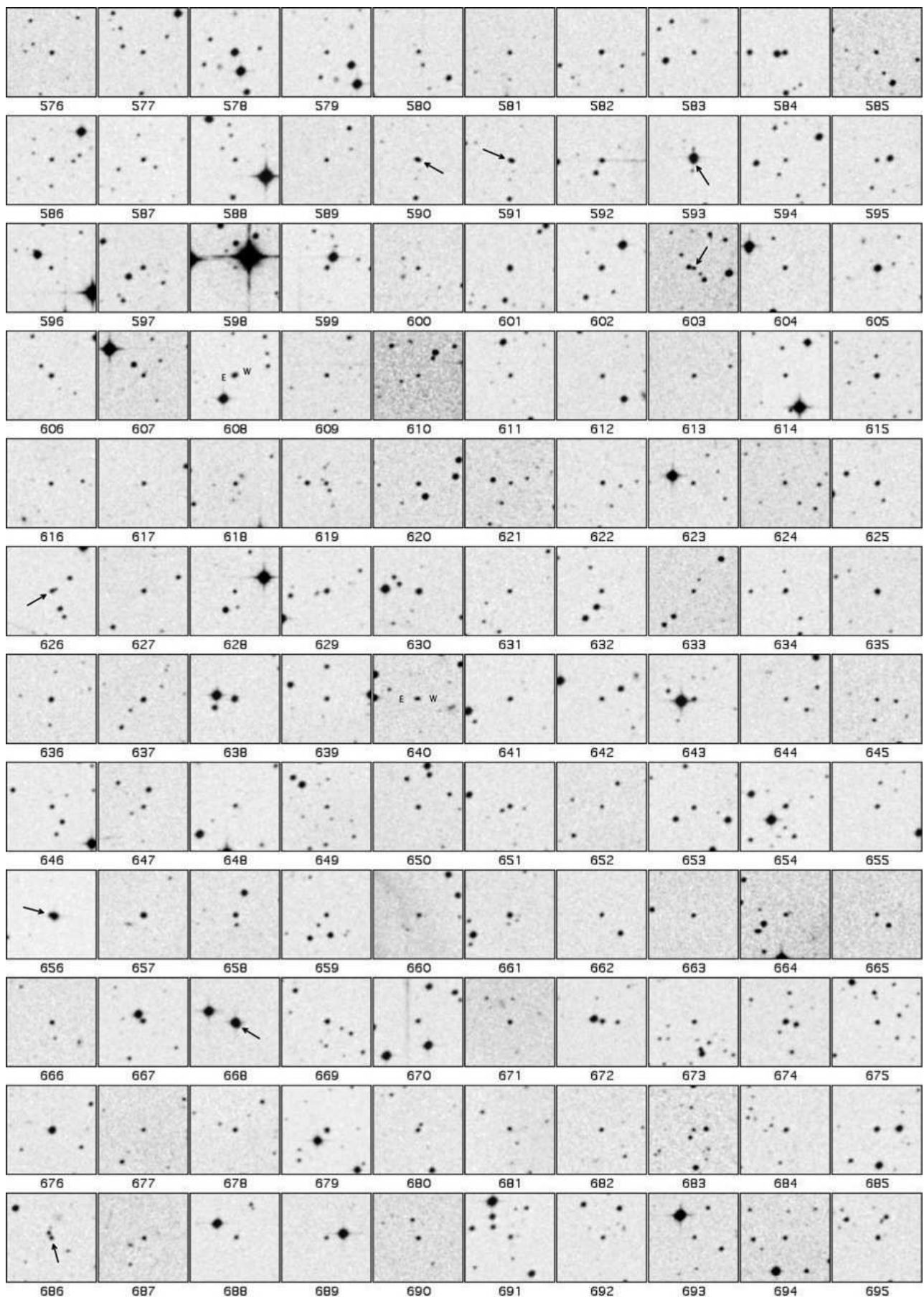}
      \caption{Finding charts, 90\arcsec\ to a side. North is up and east to the left.}
         \label{fig-8x12}
   \end{figure*}

   \begin{figure*}
   \centering
   \includegraphics[width=16.8cm]{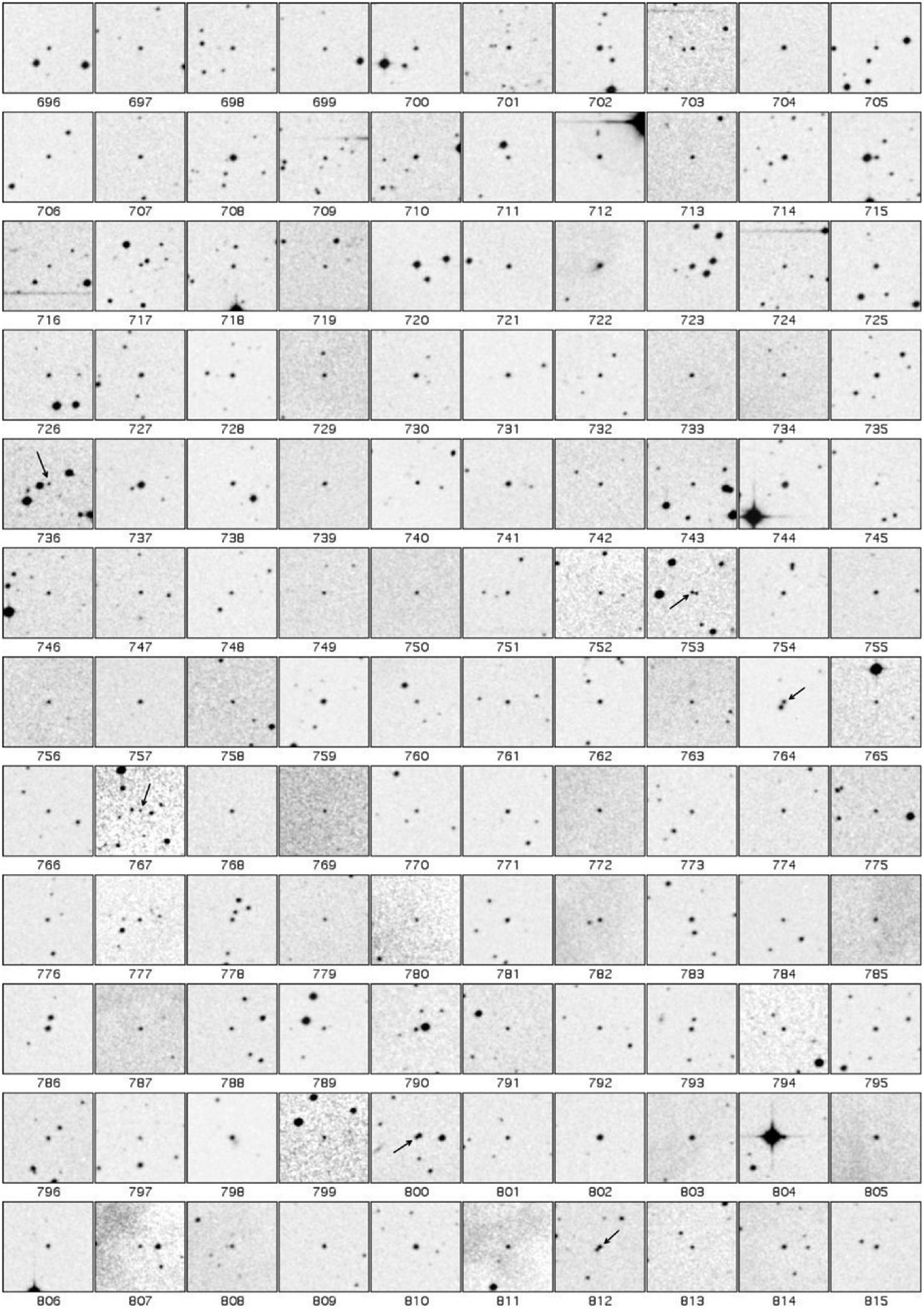}
      \caption{Finding charts, 90\arcsec\ to a side. North is up and east to the left.}
   \end{figure*}

   \begin{figure*}
   \centering
   \includegraphics[width=16.8cm]{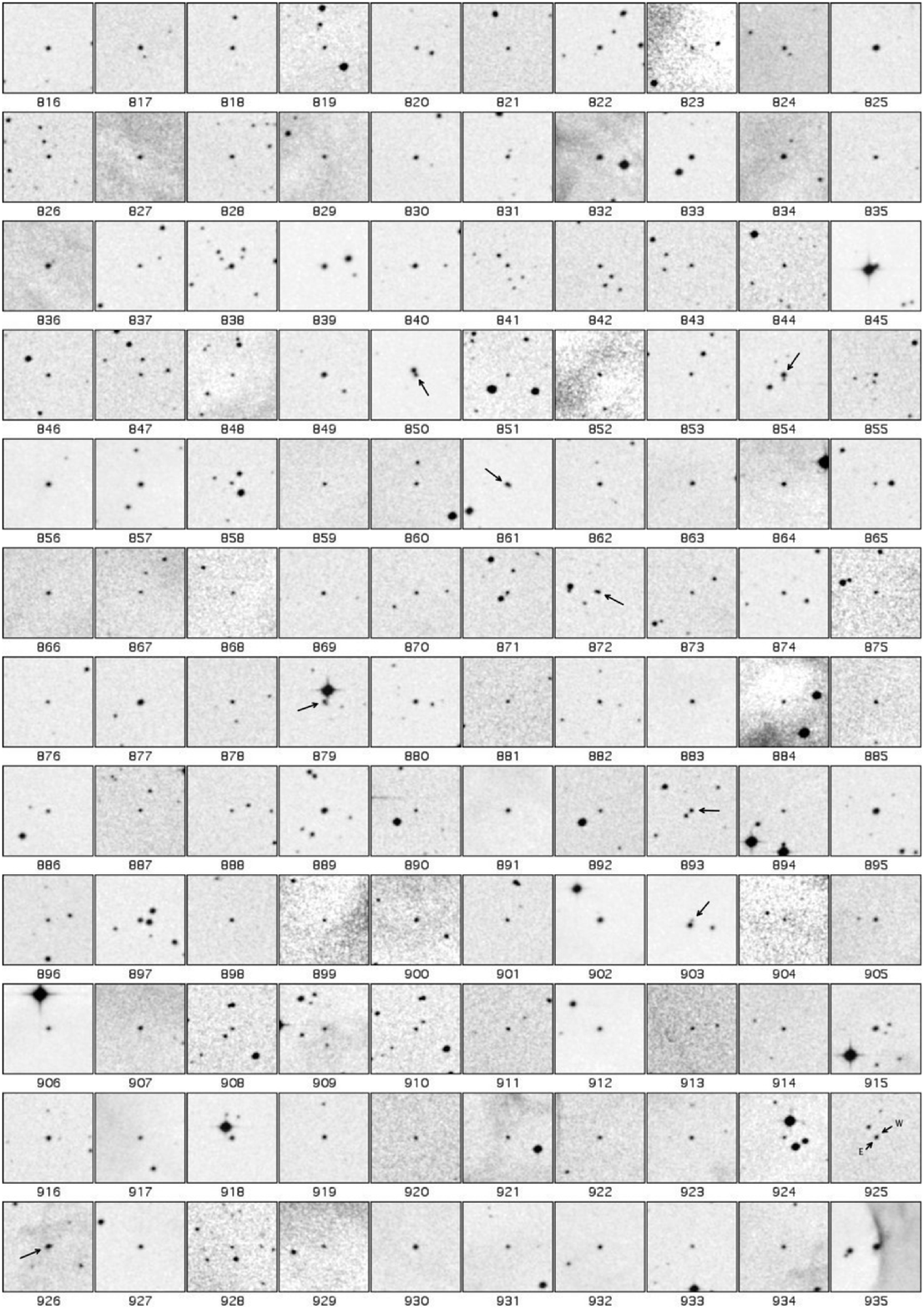}
      \caption{Finding charts, 90\arcsec\ to a side. North is up and east to the left.}
   \end{figure*}

   \begin{figure*}
   \centering
   \includegraphics[width=16.8cm]{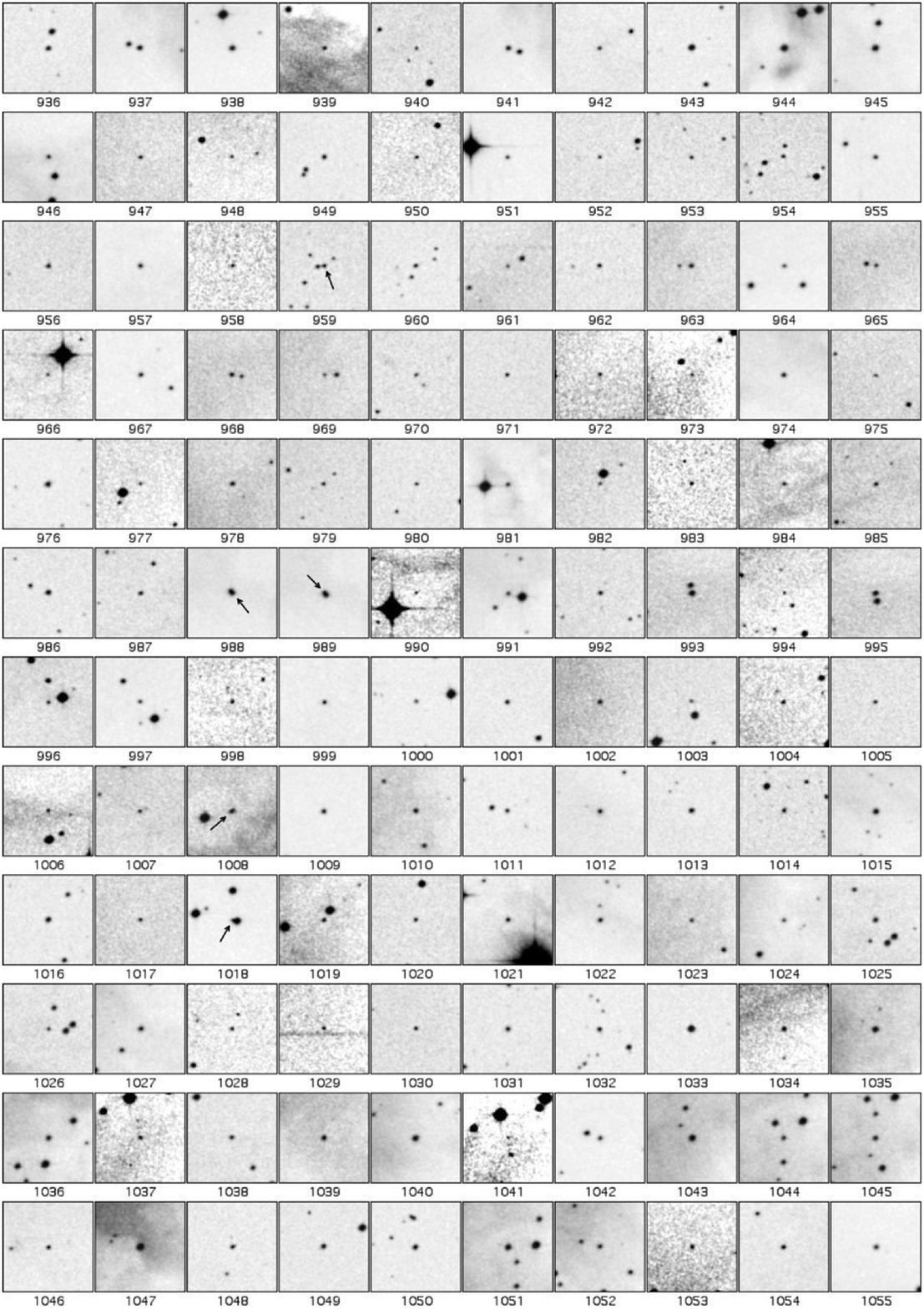}
      \caption{Finding charts, 90\arcsec\ to a side. North is up and east to the left.}
   \end{figure*}

   \begin{figure*}
   \centering
   \includegraphics[width=16.8cm]{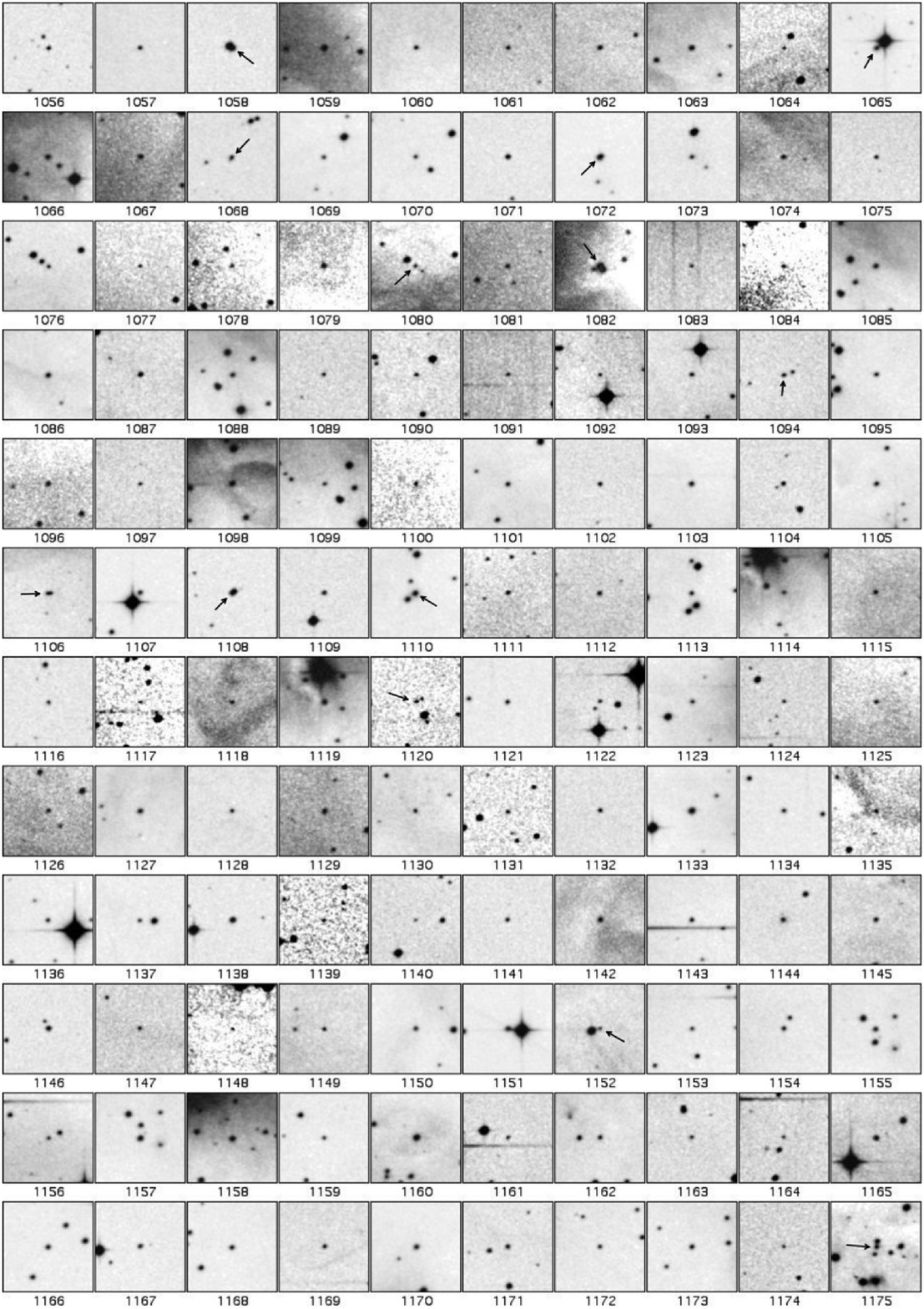}
      \caption{Finding charts, 90\arcsec\ to a side. North is up and east to the left.}
   \end{figure*}

   \begin{figure*}
   \centering
   \includegraphics[width=16.8cm]{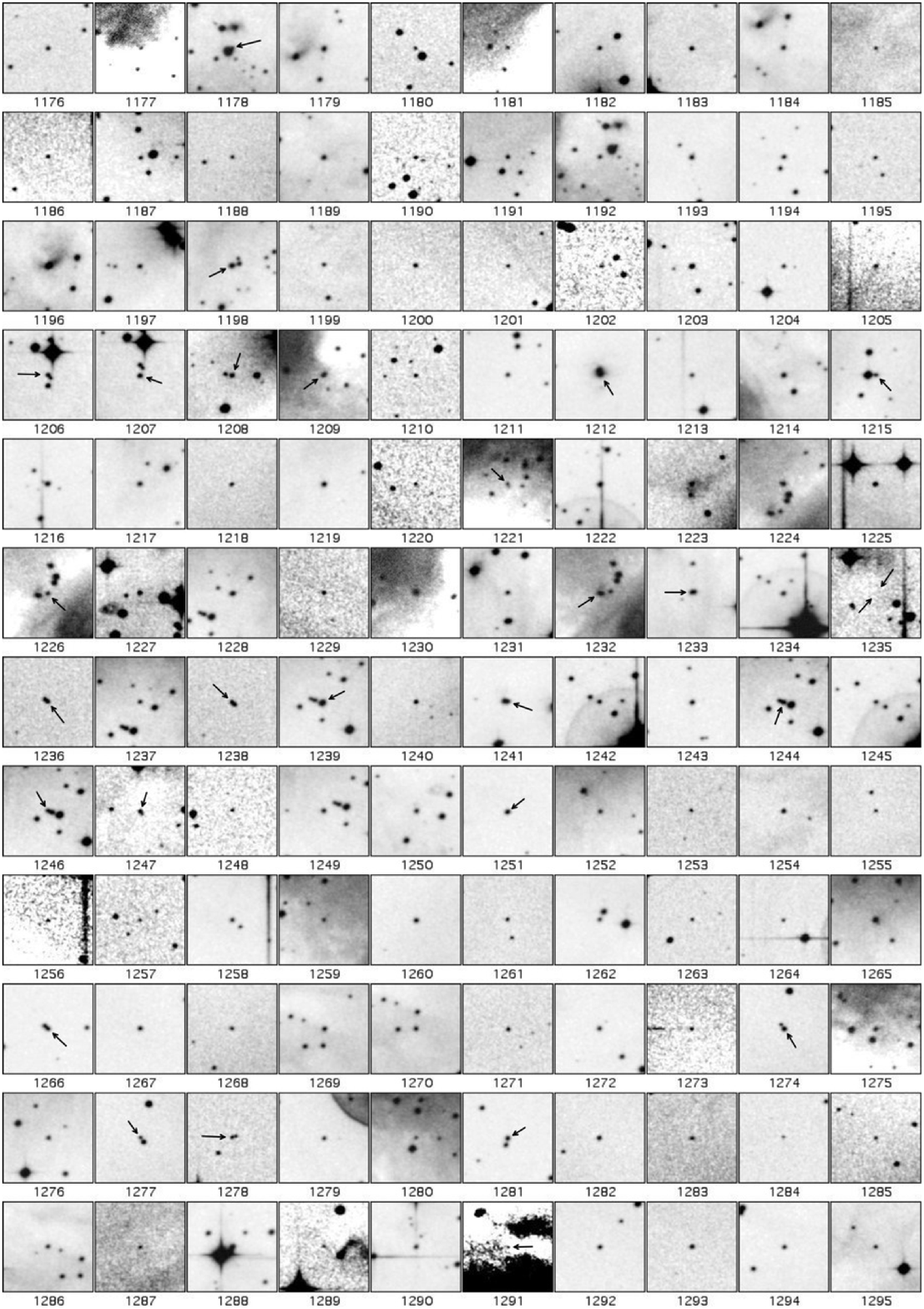}
      \caption{Finding charts, 90\arcsec\ to a side. North is up and east to the left.}
   \end{figure*}

   \begin{figure*}
   \centering
   \includegraphics[width=16.8cm]{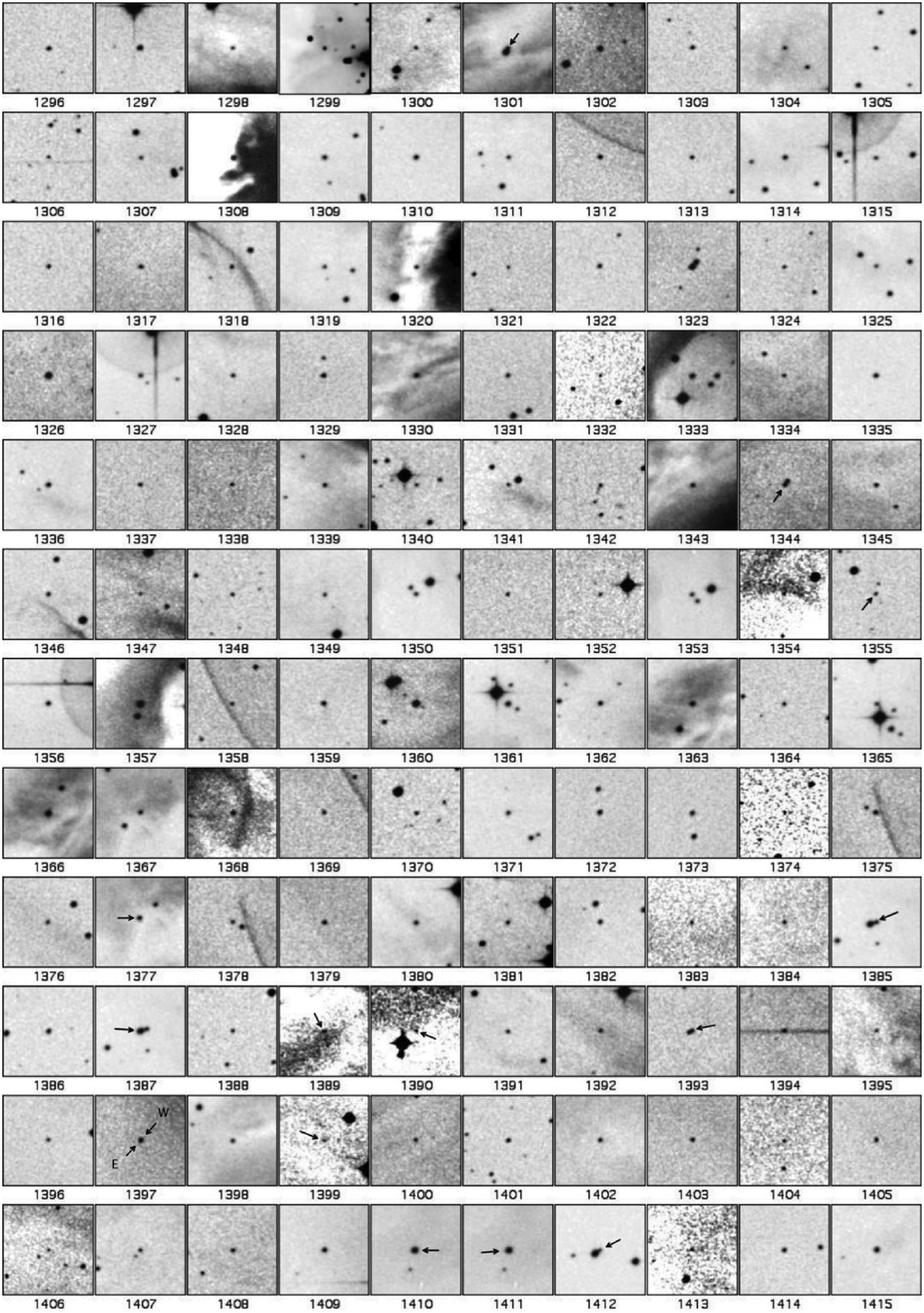}
      \caption{Finding charts, 90\arcsec\ to a side. North is up and east to the left.}
   \end{figure*}

   \begin{figure*}
   \centering
   \includegraphics[width=16.8cm]{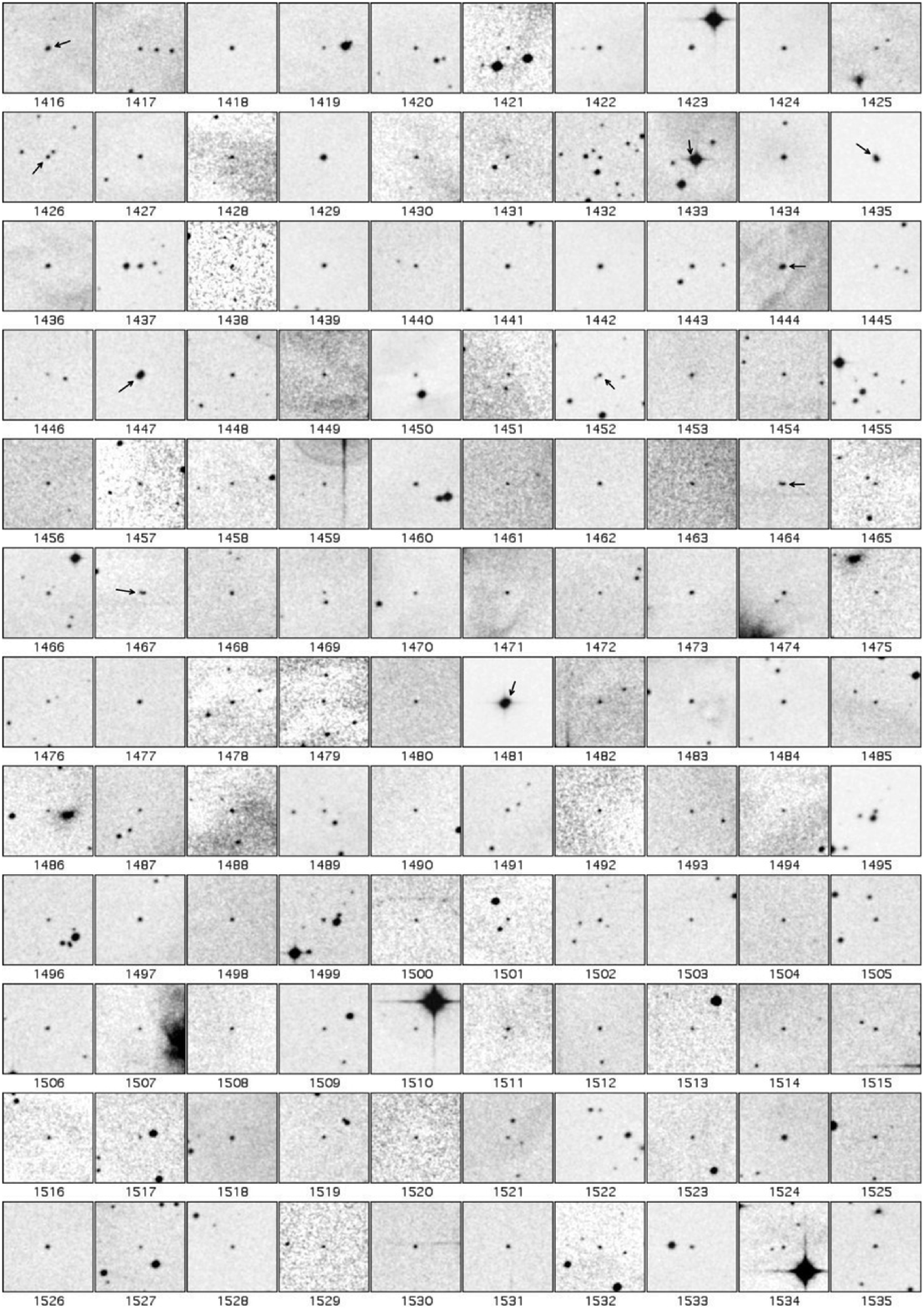}
      \caption{Finding charts, 90\arcsec\ to a side. North is up and east to the left.}
   \end{figure*}

   \begin{figure*}
   \centering
   \includegraphics[width=16.8cm]{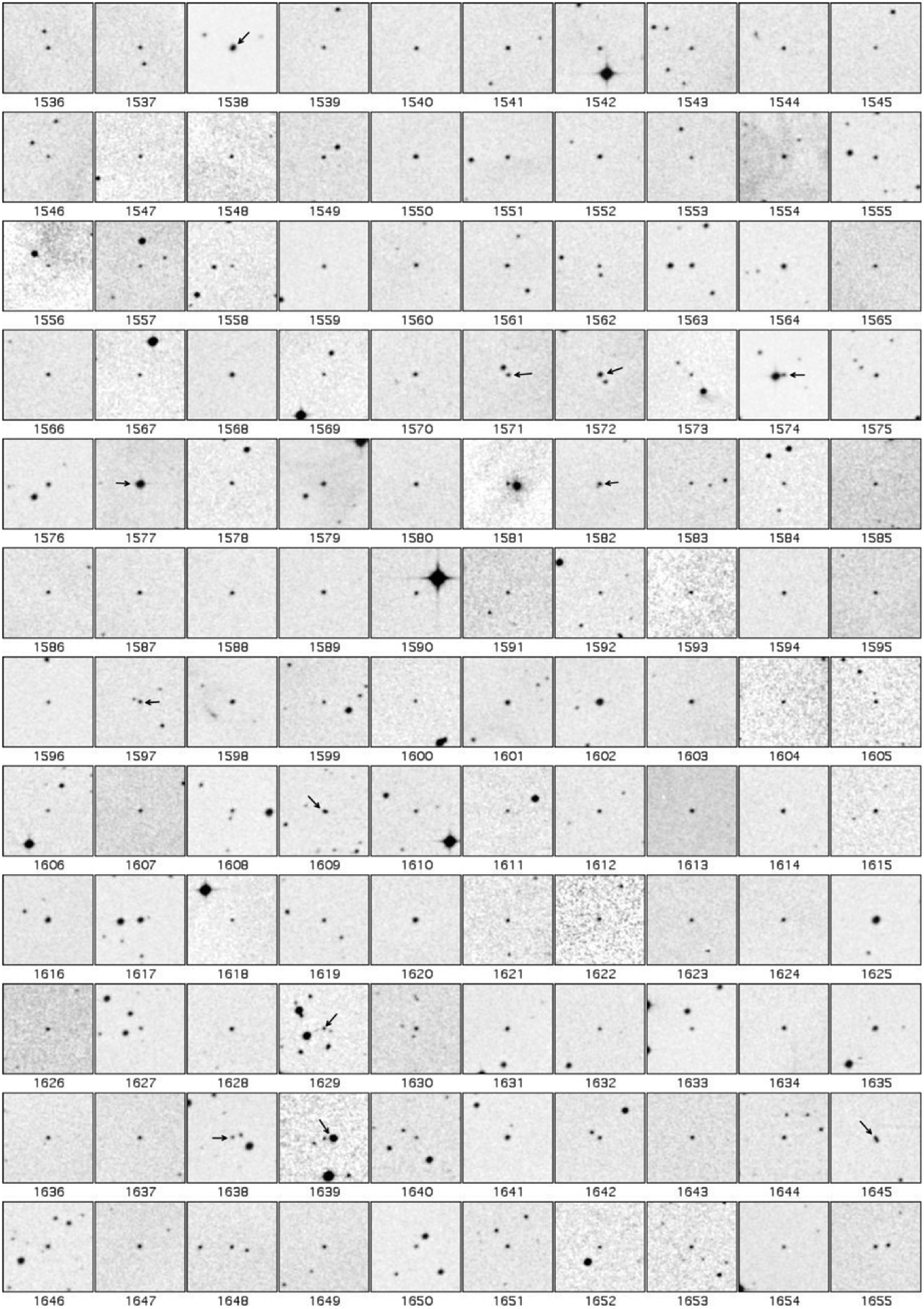}
      \caption{Finding charts, 90\arcsec\ to a side. North is up and east to the left.}
   \end{figure*}

   \begin{figure*}
   \centering
   \includegraphics[width=16.8cm]{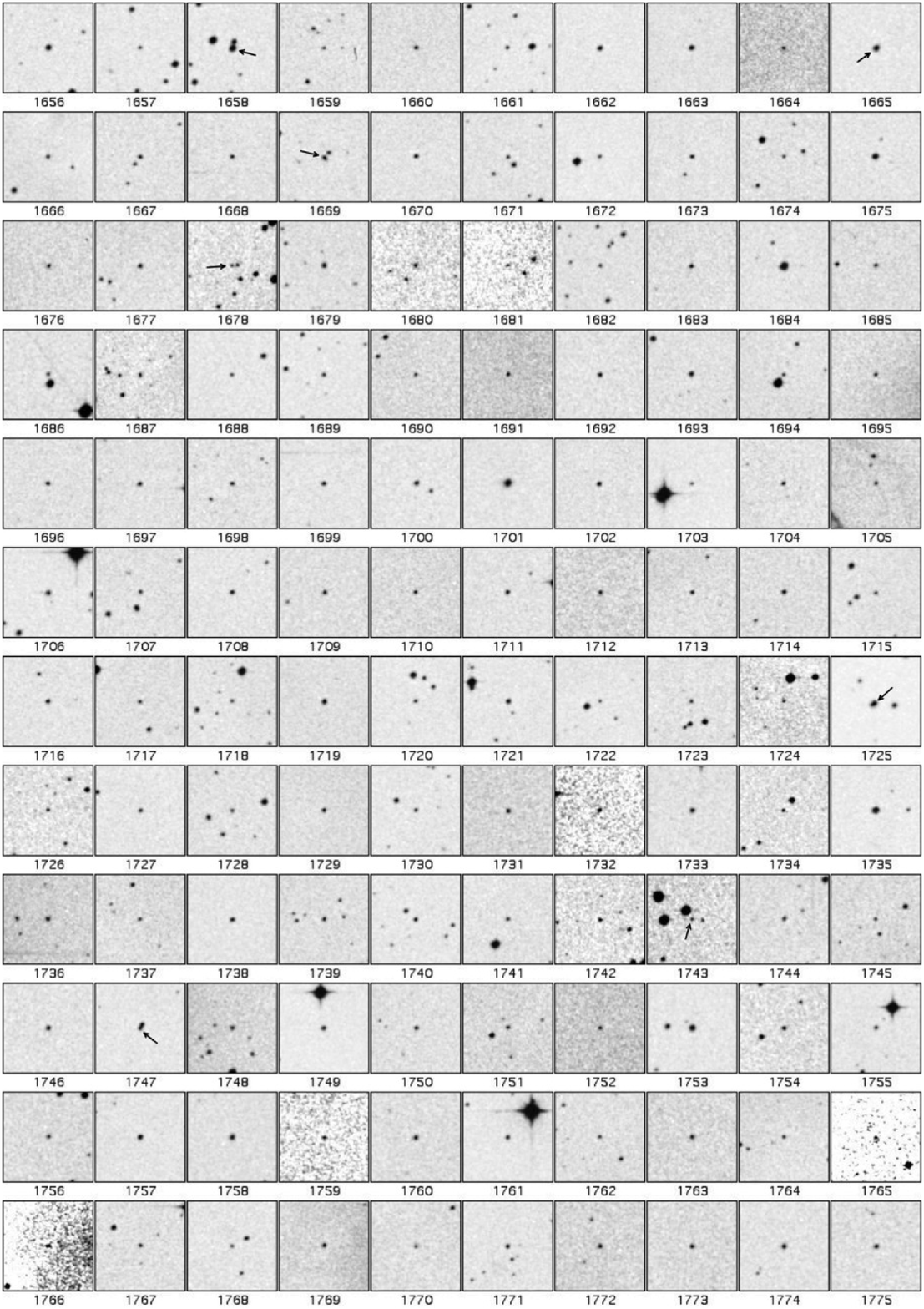}
      \caption{Finding charts, 90\arcsec\ to a side. North is up and east to the left.}
   \end{figure*}

   \begin{figure*}
   \centering
   \includegraphics[width=16.8cm]{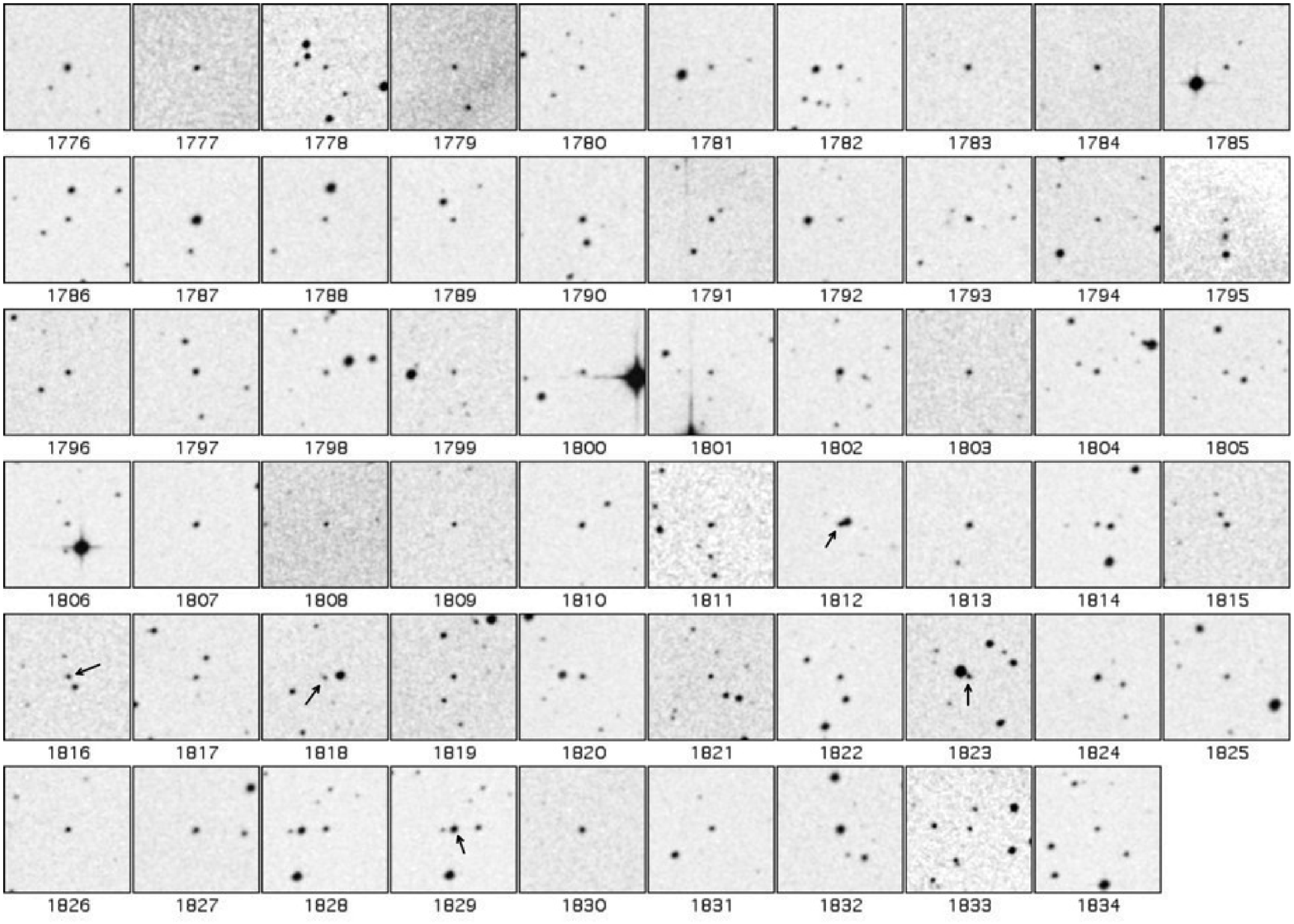}
      \caption{Finding charts, 90\arcsec\ to a side. North is up and east to the left.}
   \end{figure*}
} 


\bibliographystyle{aa}
\bibliography{Orion}

\begin{thebibliography}{36}
\expandafter\ifx\csname natexlab\endcsname\relax\def\natexlab#1{#1}\fi

\bibitem[{Bally(2008)}]{Ba08}
Bally, J. 2008, in Handbook of Star Forming Regions, Volume I: The Northern
  Sky, ed. B.~Reipurth (ASP Monograph Publications), 459

\bibitem[{{Bessell} \& {Brett}(1988)}]{BB88}
{Bessell}, M.~S. \& {Brett}, J.~M. 1988, \pasp, 100, 1134

\bibitem[{{Blaauw}(1964)}]{Blaa64}
{Blaauw}, A. 1964, \araa, 2, 213

\bibitem[{Bouy {et~al.}(2014)Bouy, Alves, Bertin, Sarro, \& Barrado}]{BoAl14}
Bouy, H., Alves, J., Bertin, E., Sarro, L., \& Barrado, D. 2014, \aap, 564, A29

\bibitem[{{Brand} \& {Wouterloot}(1992)}]{BW92}
{Brand}, J. \& {Wouterloot}, J. 1992, in ESO Scientific Reports, Vol.~11, Low
  Mass Star Formation in Southern Molecular Clouds, ed. B.~Reipurth (ESO), 1

\bibitem[{{Brice{\~n}o}(2008)}]{Br08}
{Brice{\~n}o}, C. 2008, in Handbook of Star Forming Regions, Volume I: The
  Northern Sky, ed. B.~Reipurth (ASP Monograph Publications), 838

\bibitem[{{Carpenter} {et~al.}(2001){Carpenter}, {Hillenbrand}, \&
  {Skrutskie}}]{CH01}
{Carpenter}, J.~M., {Hillenbrand}, L.~A., \& {Skrutskie}, M.~F. 2001, \aj, 121,
  3160

\bibitem[{{Da Rio} {et~al.}(2009){Da Rio}, {Robberto}, {Soderblom}, Panagia,
  L.~A.~Hillenbrand, Palla, \& K.~Stassun}]{DR09}
{Da Rio}, N., {Robberto}, M., {Soderblom}, D.~R., {et~al.} 2009, \apjs, 183,
  261

\bibitem[{Davis {et~al.}(2009)Davis, Froebrich, Stanke, \& et~al.}]{DF09}
Davis, C.~J., Froebrich, D., Stanke, T., \& et~al. 2009, \aap, 496, 153

\bibitem[{{F{\H u}r{\'e}sz} {et~al.}(2008){F{\H u}r{\'e}sz}, {Hartmann},
  {Megeath}, {Szentgyorgyi}, \& {Hamden}}]{FH08}
{F{\H u}r{\'e}sz}, G., {Hartmann}, L.~W., {Megeath}, S.~T., {Szentgyorgyi},
  A.~H., \& {Hamden}, E.~T. 2008, \apj, 676, 1109

\bibitem[{{Getman} {et~al.}(2005){Getman}, {Flaccomio}, {Broos}, {Grosso},
  {Tsujimoto}, {Townsley}, {Garmire}, {Kastner}, {Li}, {Harnden}, {Wolk},
  {Murray}, {Lada}, {Muench}, {McCaughrean}, {Meeus}, {Damiani}, {Micela},
  {Sciortino}, {Bally}, {Hillenbrand}, {Herbst}, {Preibisch}, \&
  {Feigelson}}]{GF05}
{Getman}, K.~V., {Flaccomio}, E., {Broos}, P.~S., {et~al.} 2005, \apjs, 160,
  319

\bibitem[{{Gomez} \& {Lada}(1998)}]{GoLa98}
{Gomez}, M. \& {Lada}, C.~J. 1998, \aj, 115, 1524

\bibitem[{{Haro}(1953)}]{Haro53}
{Haro}, G. 1953, \apj, 117, 73

\bibitem[{{Haro} \& {Moreno}(1953)}]{HM53}
{Haro}, G. \& {Moreno}, A. 1953, Bol. Obs. Tonantzintla Tacubaya, 1, 11

\bibitem[{{Herbig} \& {Bell}(1988)}]{HB88}
{Herbig}, G.~H. \& {Bell}, K.~R. 1988, Lick Obs. Bull., No. 1111

\bibitem[{{Herbig} \& {Kameswara Rao}(1972)}]{HR72}
{Herbig}, G.~H. \& {Kameswara Rao}, N. 1972, \apj, 174, 401

\bibitem[{{Hillenbrand}(1997)}]{Hill97}
{Hillenbrand}, L.~A. 1997, \aj, 113, 1733

\bibitem[{{Jones} \& {Walker}(1988)}]{JoWa88}
{Jones}, B.~F. \& {Walker}, M.~F. 1988, \aj, 95, 1755

\bibitem[{{Koenig} {et~al.}(2012){Koenig}, {Leisawitz}, {Benford}, {Rebull},
  {Padgett}, \& {Assef}}]{KoLe12}
{Koenig}, X.~P., {Leisawitz}, D.~T., {Benford}, D.~J., {et~al.} 2012, \apj,
  744, 130

\bibitem[{{Kogure} {et~al.}(1989){Kogure}, {Yoshida}, {Wiramihardja}, {Nakano},
  {Iwata}, \& {Ogura}}]{KYW89}
{Kogure}, T., {Yoshida}, S., {Wiramihardja}, S.~D., {et~al.} 1989, \pasj, 41,
  1195

\bibitem[{{Kukarkin}(1985)}]{Kuka85}
{Kukarkin}, B.~V. 1985, {General catalogue of variable stars, 4th ed.} (Moscow:
  Nauka Publishing House)

\bibitem[{{Maddalena} {et~al.}(1986){Maddalena}, {Morris}, {Moscowitz}, \&
  J.~\&~{Thaddeus}}]{MM86}
{Maddalena}, R.~J., {Morris}, M., {Moscowitz}, \& J.~\&~{Thaddeus}, P. 1986,
  \apj, 303, 375

\bibitem[{{Megeath} {et~al.}(2012){Megeath}, {Gutermuth}, {Muzerolle},
  {Kryukova}, {Flaherty}, {Hora}, {Allen}, {Hartmann}, {Myers}, {Pipher},
  {Stauffer}, {Young}, \& {Fazio}}]{MeGu12}
{Megeath}, S.~T., {Gutermuth}, R., {Muzerolle}, J., {et~al.} 2012, \aj, 144,
  192

\bibitem[{{Menten} {et~al.}(2007){Menten}, {Reid}, {Forbrich}, \&
  {Brunthaler}}]{MR07}
{Menten}, K.~M., {Reid}, M.~J., {Forbrich}, J., \& {Brunthaler}, A. 2007, \aap,
  474, 515

\bibitem[{{Muench} {et~al.}(2008){Muench}, {Getman}, {Hillenbrand}, \&
  {Preibisch}}]{MG08}
{Muench}, A., {Getman}, K., {Hillenbrand}, L., \& {Preibisch}, T. 2008, in
  Handbook of Star Forming Regions, Volume I: The Northern Sky, ed. B.~Reipurth
  (ASP Monograph Publications), 483

\bibitem[{{Nakano} {et~al.}(1995){Nakano}, {Wiramihardja}, \& {Kogure}}]{NW95}
{Nakano}, M., {Wiramihardja}, S.~D., \& {Kogure}, T. 1995, \pasj, 47, 889

\bibitem[{{O'Dell} {et~al.}(2008){O'Dell}, {Muench}, {Smith}, \&
  {Zapata}}]{OM08}
{O'Dell}, C.~R., {Muench}, A., {Smith}, N., \& {Zapata}, L. 2008, in Handbook
  of Star Forming Regions, Volume I: The Northern Sky, ed. B.~Reipurth (ASP
  Monograph Publications), 544

\bibitem[{{Parsamian} \& {Chavira}(1982)}]{PaCh82}
{Parsamian}, E.~S. \& {Chavira}, E. 1982, Bol. Obs. Tonantzintla Tacubaya, 3,
  69

\bibitem[{{Reipurth} {et~al.}(2004){Reipurth}, {Pettersson}, {Armond}, {Bally},
  \& {Vaz}}]{RPA04}
{Reipurth}, B., {Pettersson}, B., {Armond}, T., {Bally}, J., \& {Vaz}, L.~P.~R.
  2004, \aj, 127, 1117, (Paper~I)

\bibitem[{{Rieke} \& {Lebofsky}(1985)}]{RL85}
{Rieke}, G.~H. \& {Lebofsky}, M.~J. 1985, \apj, 288, 618

\bibitem[{{Szegedi-Elek} {et~al.}(2013){Szegedi-Elek}, {Kun}, {Reipurth},
  {P{\'a}l}, {Bal{\'a}zs}, \& {Willman}}]{SE13}
{Szegedi-Elek}, E., {Kun}, M., {Reipurth}, B., {et~al.} 2013, \apjs, 208, 28

\bibitem[{{Warren} \& {Hesser}(1978)}]{WaHe78}
{Warren}, Jr., W.~H. \& {Hesser}, J.~E. 1978, \apjs, 36, 497

\bibitem[{{Wiramihardja} {et~al.}(1991){Wiramihardja}, {Kogure}, {Yoshida},
  {Nakano}, {Ogura}, \& {Iwata}}]{WKY91}
{Wiramihardja}, S.~D., {Kogure}, T., {Yoshida}, S., {et~al.} 1991, \pasj, 43,
  27

\bibitem[{{Wiramihardja} {et~al.}(1989){Wiramihardja}, {Kogure}, {Yoshida},
  {Ogura}, \& {Nakano}}]{WKY89}
{Wiramihardja}, S.~D., {Kogure}, T., {Yoshida}, S., {Ogura}, K., \& {Nakano},
  M. 1989, \pasj, 41, 155

\bibitem[{{Wiramihardja} {et~al.}(1993){Wiramihardja}, {Kogure}, {Yoshida},
  {Ogura}, \& {Nakano}}]{WKY93}
{Wiramihardja}, S.~D., {Kogure}, T., {Yoshida}, S., {Ogura}, K., \& {Nakano},
  M. 1993, \pasj, 45, 643

\bibitem[{{Wouterloot} \& {Brand}(1992)}]{WB92}
{Wouterloot}, J.~G.~A. \& {Brand}, J. 1992, \aap, 265, 144

\end{thebibliography}



\newpage

\onecolumn
\onllongtab{
{\scriptsize
\tabcolsep 5pt
\centering

\begin{longtable}{lllllllcccccccccl}
\caption{\label{tbl3} H$\alpha$ emission-line stars in M 42$^*$}\\
\hline\hline
ESOH$\alpha$ & GCVS & KisoH$\alpha$\tablefootmark{a} & HBC\tablefootmark{b} & Haro\tablefootmark{c} & PaCh & Other\tablefootmark{d}  &
$\alpha$(2000)\tablefootmark{e} & $\delta$(2000)$^e$ & H$\alpha$\tablefootmark{f} &  
$m_{J}$\tablefootmark{g} & $m_{F}^\mathrm{g}$ & $m_{N}^\mathrm{g}$ & $J$\tablefootmark{h} & $H^\mathrm{h}$ & $K_s^\mathrm{h}$&Notes
\tablefootmark{i} \\
\hline
\endfirsthead
\caption{continued.}\\
\hline\hline
ESOH$\alpha$ & GCVS & KisoH$\alpha$ & HBC & Haro & PaCh &Other  &
$\alpha$(2000) & $\delta$(2000) & H$\alpha$ &  
$m_{J}$ & $m_{F}$ & $m_{N}$ & $J$ & $H$ & $K_{s}$&Notes \\
\hline
\endhead
\hline
\endfoot
576	&		&		&		&		&		&		&	05	24	45.32	&	$-$04	03	40.3	&	1-2	&	18.65	&	16.94	&	14.90	&	13.49	&	12.84	&	12.56	&		\\
577	&		&		&		&		&		&		&	05	24	49.82	&	$-$03	59	02.4	&	1	&	18.86	&	16.83	&	14.75	&	13.66	&	12.98	&	12.73	&		\\
578	&		&		&		&		&		&		&	05	25	04.40	&	$-$06	50	16.1	&	2	&	15.93	&	14.26	&	12.94	&	11.93	&	11.26	&	11.04	&		\\
579	&		&		&		&		&		&		&	05	25	06.06	&	$-$06	50	03.4	&	2	&	18.82	&	17.01	&	14.92	&	13.75	&	13.08	&	12.84	&		\\
580	&		&		&		&		&		&		&	05	25	19.24	&	$-$07	21	33.3	&	3	&	18.87	&	16.84	&	14.55	&	13.40	&	12.74	&	12.45	&		\\
581	&		&		&		&		&		&		&	05	25	20.61	&	$-$06	05	06.1	&	4	&	18.48	&	16.91	&	14.80	&	13.19	&	12.55	&	12.20	&		\\
582	&		&		&		&		&		&		&	05	25	20.93	&	$-$05	19	30.3	&	1	&	17.64	&	16.24	&	14.14	&	12.82	&	12.21	&	11.94	&		\\
583	&		&		&		&		&		&		&	05	25	26.16	&	$-$03	43	30.9	&	2-3	&	18.79	&	16.87	&	14.62	&	13.32	&	12.66	&	12.37	&		\\
\hline
\end{longtable}
\tablefoot{
\tablefoottext{a}{Kiso H$\alpha$ catalogue from \citet{WKY93}}
\tablefoottext{b}{\citet{HB88} catalogue.}
\tablefoottext{c}{\citet{Haro53} catalogue: Haro\,4-1 to 4-255, \citet{PaCh82}: Haro4-256 to 4-495, \citet{HM53}: Haro5-1 to 5-98.}
\tablefoottext{d}{An asterisk means a star also listed in \citet{SE13}}
\tablefoottext{e}{Positions extracted from the 2MASS All-Sky Catalog.}
\tablefoottext{f}{The H$\alpha$ strength is defined so 1 is weak emission against 
a strong continuum and 5 is strong emission against a weak or invisible continuum. Hyphenated 
values may represent either variability and/or uncertainty in the estimate. A '+' indicates resolved spectra in unresolved DSS image}
\tablefoottext{g}{The magnitudes $m_{J}$, $m_{F}$, and $m_{N}$ are from the blue
(IIIaJ emulsion), red (IIIaF emulsion), and infrared (IV-N emulsion)
digitised sky survey, extracted from USNO-B catalogue or from GSC 2.2.}
\tablefoottext{h}{$JHK_s$ magnitudes extracted from the 2MASS All-Sky Catalogue.}\\
\tablefoottext{i}{
Notes to individual stars:\\
1 - Double with 591.
2 - Unresolved photometry.
3 - Double with 590.
4 - 3\arcsec\ S of Kiso 75-87. Photometry from SDSS.
5 - W component of close pair.
6 - Unresolved binary 608W and 608E. H$\alpha$ emission seen in both components, strength 2 and 3, resp.
7 - Unresolved binary 640W and 640E. H$\alpha$ emission seen in both components, strength 4 and 1, resp.
8 - NE component of close pair.
9 - W component of unresolved pair.
10 - In small nebulosity.\\
11 - E component of close pair.
12 - Binary, emission in W, fainter component.
13 - SE component of close pair.
14 - Photometry includes faint component to W. 
15 - Double with Kiso 76-91. 
16 - Double with PaCh 115.
17 - This object is not V750 Ori. Misidentified in the Simbad database.
18 - Double with 989.
19 - Double with 988.
20 - Chanal's object.\\
21 - Double with 1018. $m_{J}m_{F}m_{N}$ photometry in common.
22 - Double with Kiso 76-114. $m_{J}m_{F}m_{N}$ photometry in common.
23 - $m_{J}m_{F}m_{N}$ in common with Kiso 76-118.
24 - $m_{J}m_{F}m_{N}$ in common with 1058.
25 - NW component of close pair.
26 - Double with 1072.
27 - Double with Kiso 76-124.
28 - I from \citet{JoWa88}.
29 - Stellar.
30 - Unresolved double.\\
31 - SW component of close pair.
32 - Close to 1233.
33 - Close to Kiso 76-162.
34 - Very faint continuum, no photometry.
35 - $m_{J}m_{F}m_{N}$ unresolved from 1238.
36 - $m_{J}m_{F}m_{N}$ unresolved from 1236.
37 - 1\farcs3 NW of V417 Ori.
38 - Photometry unresolved from 1239.
39 - $m_{J}m_{F}m_{N}$ photometry unresolved.
40 - Double with Haro 4-164.\\
41 - Double with 1251.
42 - Double with Kiso 76-188.
43 - Double with 1301.
44 - Double with 1411.
45 - Double with 1410.
46 - Double with Kiso 76-231.
47 - Double with 1412.
48 - Photometry contaminated by faint component to SE.
49 - SW component of binary AX Ori.
50 - NE component of brighter AX Ori.\\
51 - Double with Kiso 76-243.
52 - Double with 1444.
53 - Double with 1467.
54 - Double with 1464.
55 - In L1641 N.
56 - Cohen-Schwartz's star.
57 - NW of and near brighter PR Ori. No photometry.
58 - Double with 1538.
59 - NW comp to V846.
60 - Double with 1645.\\
61 - Double with Haro 4-439.
62 - Sugano's star, Abastumani 24.
63 - Has a faint comp to NW.
64 - On rim of HH 449.
65 - In nebulosity.
66 - Also given as Haro 4-461 by \citet{PaCh82}.
67 - Double with Kiso 76-368.
68 - Double with 1725.
69 - has IR comp to SW.
} 
} 
}  
}  
\end{document}